\documentclass[a4paper]{raa}
\usepackage{graphicx,times}
\usepackage{natbib}
\usepackage{amssymb,amsmath}
\usepackage{ulem}
\usepackage{color}
\usepackage[colorlinks, citecolor=blue]{hyperref}
\usepackage{longtable}

\begin{document}

    \title{The Data Processing of the LAMOST Medium-Resolution Spectral Survey of Galactic Nebulae (LAMOST \textbf{\textit{MRS-N Pipeline}})}

 	\volnopage{ {\bf 2012} Vol.\ {\bf X} No. {\bf XX}, 000--000}
   	\setcounter{page}{1}

	\author{Chao-Jian Wu\inst{1,2}, Hong Wu\inst{1,2}, 
			Wei Zhang\inst{1,2}, Yao Li\inst{2,3}, Juan-Juan Ren\inst{2,4}, 
			Jian-Jun Chen\inst{1,2}, Chih-Hao Hsia\inst{5}, 
			Yu-Zhong Wu\inst{1,2}, Hui Zhu\inst{2}, 
			Bin Li\inst{6,7}, Yong-Hui Hou\inst{3,8}, 
		}

	\institute{CAS Key Laboratory of Optical Astronomy, National Astronomical Observatories, Chinese Academy of Sciences, Beijing 100101, China; {\it chjwu@bao.ac.cn}\\
	\and
				National Astronomical Observatories, Chinese Academy of Sciences, 20A Datun Road, Chaoyang District, Beijing 100101, China\\
	\and
                School of Astronomy and Space Science, University of Chinese Academy of Sciences, Beijing 100049, China;
	\and
				CAS Key Laboratory of Space Astronomy and Technology, National Astronomical Observatories, Chinese Academy of Sciences, Beijing 100101, China\\ 
	\and
				State Key Laboratory of Lunar and Planetary Sciences, Macau University of Science and Technology, Taipa, Macau, China \\
	\and
				Purple Mountain Observatory, Chinese Academy of Sciences, Nanjing 210008, People's Republic of China\\
	\and
				University of Science and Technology of China, Hefei 230026, People's Republic of China\\		
	\and
				Nanjing Institute of Astronomical Optics, \& Technology, National Astronomical Observatories, Chinese Academy of Sciences, Nanjing 210042, China \\
	\vs \no
	{\small Received ---; accepted ---}
	}

\abstract{
	The Large sky Area Multi-Object Fiber Spectroscopic Telescope (LAMOST) medium-resolution spectral survey of Galactic Nebulae (MRS-N) has conducted for more than three years since Sep. 2018 and observed more than 190 thousands nebular spectra and 20 thousands stellar spectra. However, there is not yet a data processing pipeline for nebular spectra. To significantly improve the accuracy of nebulae classification and their physical parameters, we developed the \textbf{\textit{MRS-N Pipeline}}. This article presented in detail each data processing step of the \textbf{\textit{MRS-N Pipeline}}, such as removing cosmic rays, merging single exposure, fitting sky light emission lines, wavelength recalibration, subtracting skylight, measuring nebular parameters, creating catalogs and packing spectra. Finally, a description of the data products, including nebular spectra files and parameter catalogs, is provided.
     	\keywords{surveys --- catalogs --- methods: data analysis --- ISM: general}
}

\authorrunning{Ch.J. Wu et al.}
\titlerunning{The \textbf{\textit{MRS-N Pipeline}}}
\maketitle

\section{Introduction}\label{s:intro}

The LAMOST MRS-N, \citep{wu2020, RAA4768}, as a sub-project of the Medium-Resolution Spectral Survey (MRS, \citep{liu2020}), mainly relies on the LAMOST (also known as Guoshoujing Telescope), which is the first astronomical large-scale scientific facility of China \citep{cui2012, zhao2012, 1996ApOpt..35.5155W, 2004ChJAA...4....1S} and has achieved great success in many astronomical fields \citep{liu2020, 2014ApJ...788L..37G, 2015ApJ...807....4L, 2016NatCo...711058K, 2016ApJ...818..202L, 2017RAA....17...10W, 2018RAA....18..111R, 2019MNRAS.490..550L, 2019Natur.575..618L, 2019ApJ...872L..20G,  2016RAA....16..102W, wu2020, 2015ApJ...809..145T, 2015RAA....15.1209X, 2016MNRAS.463.2623H, 2017RAA....17...96L, 2017ApJ...835L..18L, 2017RAA....17..114T, 2017MNRAS.470.2949W, 2018ApJS..238...16L, 2018ApJ...865L..19T, 2018MNRAS.478.3367W, 2018MNRAS.473.1244X, 2018MNRAS.475.1093Y, 2018ApJ...868..105Z, 2019ApJ...874..138L, 2019ApJ...871..184T, 2019ApJ...877L...7W, 2019MNRAS.482.2189W, 2016RAA....16...43S, 2010RAA....10..737W, 2010RAA....10..745W, 2019ApJ...884L...7H, 2021MNRAS.506.6117W}, to observe the nebulae (including the \ion{H}{ii} regions, Herbig-Haro objects, supernova remnants, planetary nebulae) in the northern Galactic Plane (GP). 
From October 2018 to now, MRS-N has been conducted for three years and obtained  more than 190 thousand medium resolution spectra of Galactic nebulae. It is one of the largest nebular survey in the world \citep{RAA4768}.  

Several pipelines have been developed to reduce the LAMOST raw data. LAMOST 2D Pipeline \citep{2015RAA....15.1095L} is used to extract spectra from CCD images, like subtracting dark and bias, correcting flat field, extracting spectra, calibrating wavelength, subtracting sky light, merging spectra, etc., and do flux calibration, while 1D Pipeline \citep{2015RAA....15.1095L}, which is used after 2D Pipeline, works on the classification and parameter measurement of stars, galaxies or quasars (QSOs). \cite{2015MNRAS.448..822X} introduced the LAMOST stellar parameter pipeline at Peking University (LSP3), which is parallel to the LAMOST 1D Pipeline. The main function of LSP3 is to determinate the radial velocities (RVs) and stellar atmospheric parameters, such as $T_\mathrm{eff}$, log $g$ and [Fe/H]. However, the LAMOST 1D Pipeline and LSP3 are mainly used to measure stellar parameters. They are both not suitable for the data of MRS-N. A MRS-N spectrum contains dozens of strong sky light emission lines, nebular emission lines and faint continuum. Unlike the stellar spectra, it is difficult to reduce the sky light effectively  for MRS-N spectra. Especially in a large scale nebular region, it is very difficult to find the spectrum of pure sky light. That is to say, the sky light of MRS-N data cannot be reduced with the regular method in the LAMOST 2D Pipeline. Moreover, the measurement of nebular parameters is also different from the measurement of stellar spectra. Specific algorithms are required. Then we developed a new pipeline, which was named \textbf{\textit{MRS-N Pipeline}} and should be used by combining with the LAMOST 2D Pipeline, only for the MRS-N data. 

In this paper, we first introduce the MRS-N observation of past three years in Section \ref{s:obs}. Section \ref{s:pip} describes merging single exposure, wavelength calibration, subtracting sky light and measuring nebular parameters in detail. The data products are presented in Section \ref{s:dp}. Finally, Section \ref{s:summary} gives a brief summary.

\section{Observation}
\label{s:obs}

From October 2018 to now, we have completed more than three years of MRS-N observation. The observations in the three years are from Oct. 2018 to Jan. 2019, Nov. 2019 to Mar. 2020 and Oct. 2020 to Mar. 2021, respectively. As described in \cite{RAA4768}, MRS-N observation should be carried on at moonless time. Due to the weather and some irresistible reasons, only 9 days were suitable for observing and finally 12 plates were finished in the first year. 
We have optimized the observation strategy comparing with the first year, then 31 plates were observed in the second year. Among the 31 plates, 20 observed plates were covered into an united area included Rosette Nebula and NGC2264 (\textit{\textbf{Ros}} area; \cite{RAA4768}), 2 plates were covered in the Westerhout 5 area (\textit{\textbf{West}}) and the left 9 were observed to cover the GP area. In the second year, we finished a complete specific area, the \textit{\textbf{Ros}} area. In the third year, mainly due to the weather, the number of observed areas decreased. Only 11 days were suitable for observing and finally 17 plates were finished in the third year. 

\begin{figure}
    \centering
    \includegraphics[width=\textwidth]{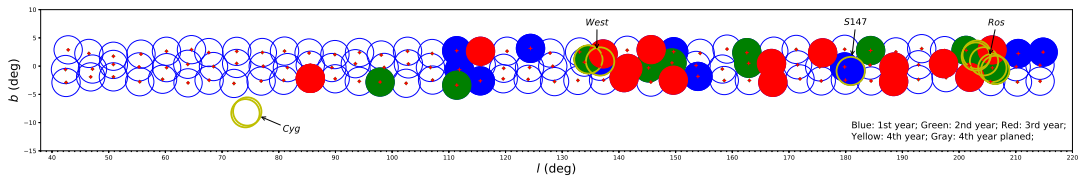}
    \caption{MRS-N coverage of past three years' observation. The x-axis represents the Galactic longitude and the y-axis represents the Galactic latitude. The blue circles indicate observation of the first year, green and red circles indicate observations of the second and the third year, respectively. The yellow circles show the 4 specific areas.}
    \label{fig:cover}
\end{figure}

Figure \ref{fig:cover} shows the three years' coverage of MRS-N. The blue circles indicate observations of the first year, the green circles indicate the second year's observations and the red circles indicate the third year's observations. The yellow circles show the 4 specific areas with names marked. All the observed MRS-N data will be processed by using the pipeline described below. 

\section{Mehtodology}
\label{s:pip}

\begin{figure}
    \centering
    \includegraphics[width=0.9\textwidth]{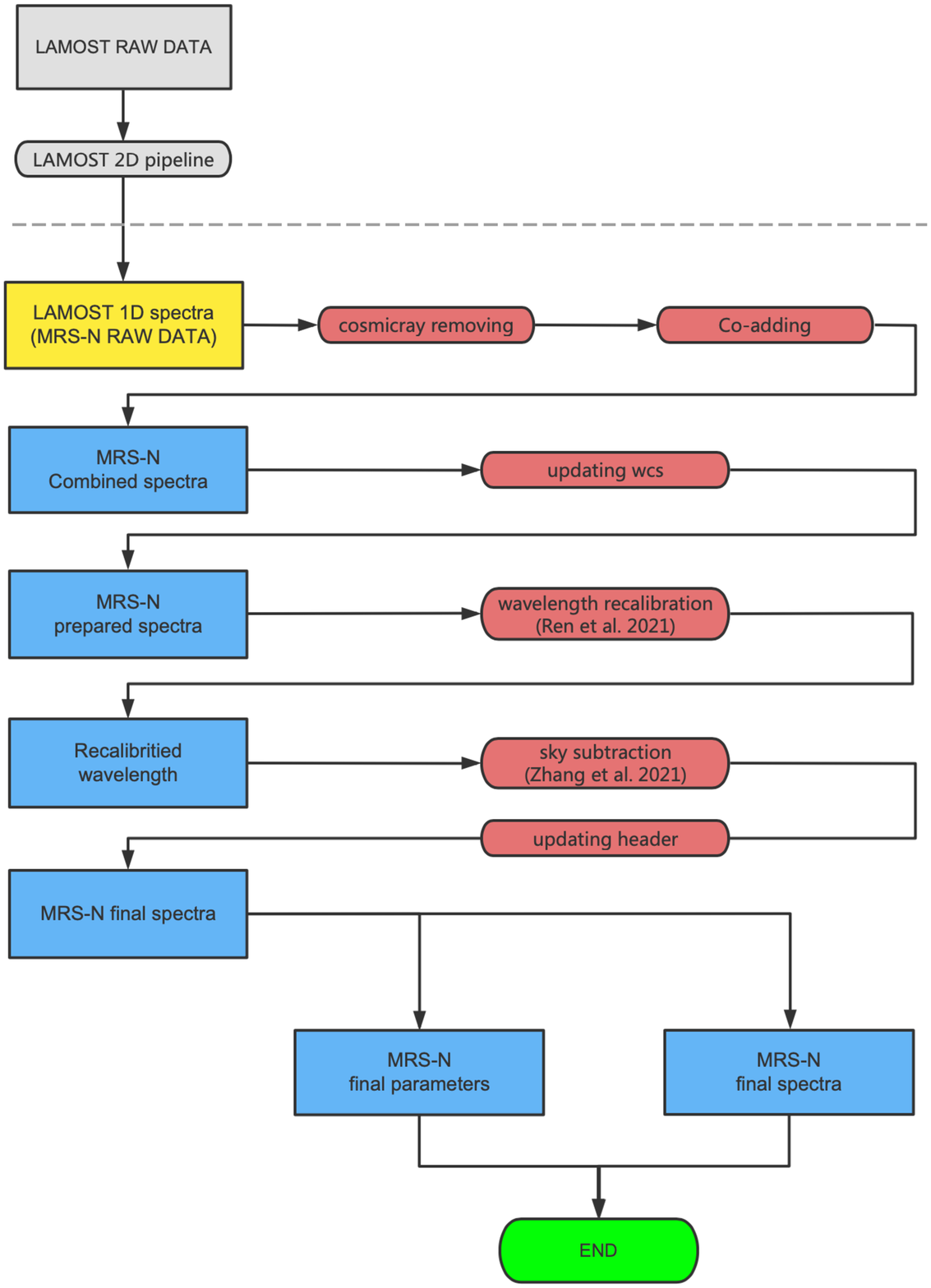}
    \caption{The \textbf{\textit{MRS-N Pipeline}} flowchart.}
    \label{fig:flow}
\end{figure}

As described in \cite{RAA4768}, the data processing of MRS-N is different from the stellar spectra reduction. The LAMOST 1D pipeline \citep{2012RAA....12.1243L, 2015RAA....15.1095L} and LSP3 \citep{2015MNRAS.448..822X} are not suitable for MRS-N data. So we developed the \textbf{\textit{MRS-N Pipeline}}. The \textbf{\textit{MRS-N Pipeline}} includes removing cosmic rays, merging single exposure, fitting skylight emission lines \citep{2021RAA....21...51R}, subtracting skylight  \citep{2021arXiv210808021Z}, wavelength recalibration, parameters measurement of nebular emission lines, creating catalogs and packing spectra. Figure \ref{fig:flow} illustrates a flowchart of \textbf{\textit{MRS-N Pipeline}}. 

In Figure \ref{fig:flow}, the two gray boxes above the dashed line, which does not belong to the \textbf{\textit{MRS-N Pipeline}}, are the steps from LAMOST official 2D pipeline. The LAMOST 1D MRS-N spectra, reduced by the LAMOST 2D pipeline without subtracting sky light and correcting flat fields, are called the MRS-N raw data in this work (yellow box in the flowchart). After removing cosmic rays (or the false sharp emission lines) and merging for the MRS-N raw data, the MRS-N combined spectra are obtained. For those MRS-N combined spectra without equatorial coordinates, the coordinates can be calculated by using the pixel coordinates in the 2D images.  \textbf{\textit{MRS-N Pipeline}} provides a linear relationship through the known equatorial coordinates and known pixel positions of the spectra. By using the relationship, the coordinates of those spectra with only pixel positions can be estimated. The purpose of this step is to improve the utilization of fibers. After this step, we get the MRS-N prepared spectra. By measuring the sky lines of the MRS-N prepared spectra, the wavelength can be recalibrated with the method of \cite{2021RAA....21...51R}. After subtracting the sky background for the wavelength recalibrated spectra, the MRS-N final spectra will be obtained. And then, some information generated during data reduction is added to fits (Flexible Image Transport System) header. The last step of \textbf{\textit{MRS-N Pipeline}} is nebular parameters measurement.

\subsection{Merging single exposure}

A MRS-N plate contains 3 single exposure spectra, each of which is exposed for 900s. The first step of \textbf{\textit{MRS-N Pipeline}} is to coadd the 3 single exposure spectra. 
The main purpose of merging spectra is to improve the signal-to-noise (S/N) ratio. The common merging methods are median merging and mean (sum) merging. The advantage of median is that it can effectively remove the cosmic rays, but the disadvantage is that the S/N is lower than the mean measurement. 
Mean merging gives a narrower (less noisy) distribution than merging by median, though both substantially reduced the width of the distribution. The conclusion so far is that combining by averaging is mildly preferable to combining by median. Computationally, the mean is also faster to compute than the median.

\begin{figure}
    \centering
    \includegraphics[width=\textwidth]{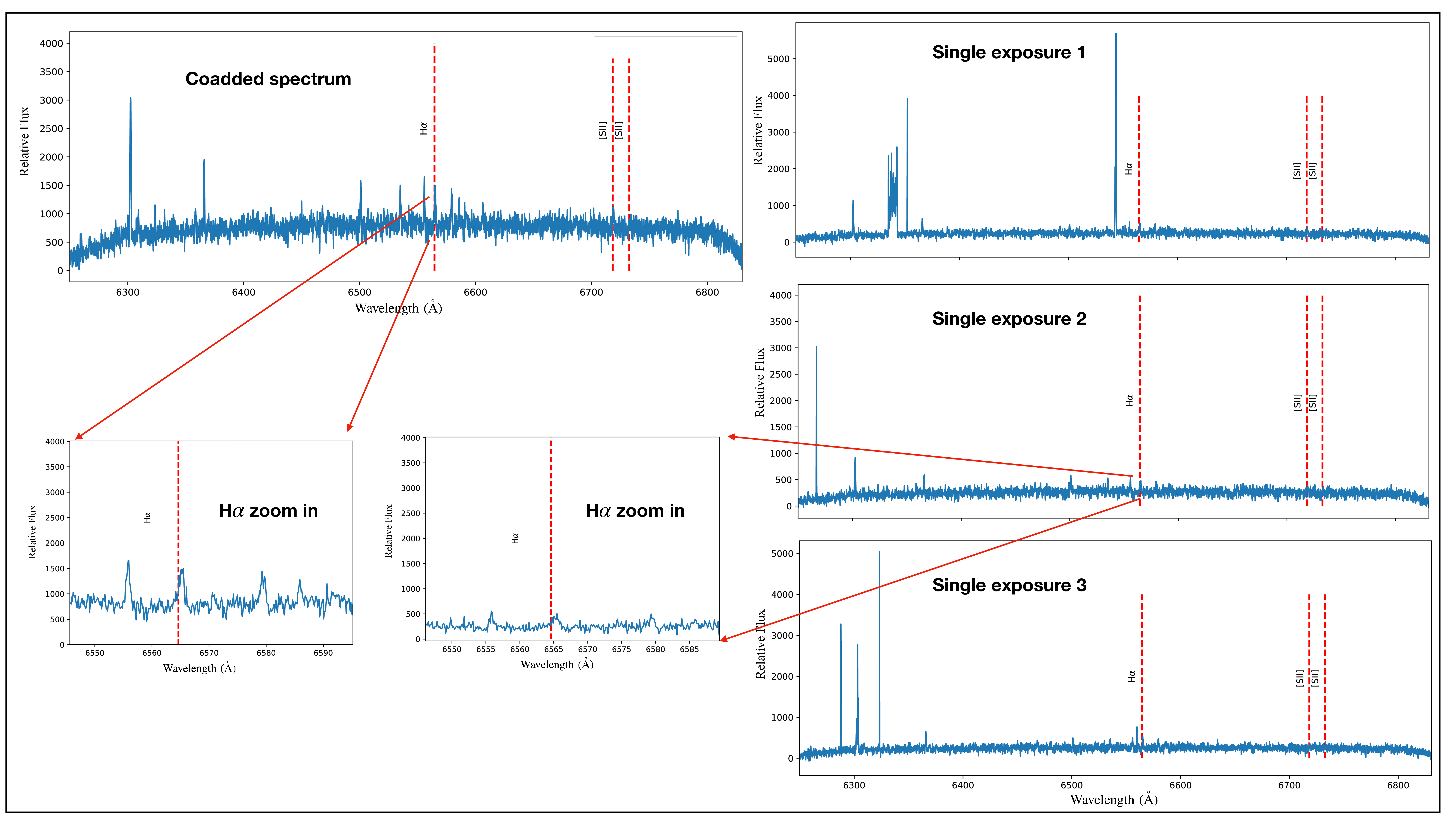}
    \caption{An example of merging single exposure. From the two lower left panels, it is clear that the S/N of merged $\mathrm{H\alpha}$ (left) is improved significantly. The red dotted lines in each panel represent nebular emission lines. It shows that the false sharp emission lines in Single exposure 1 (red ellipse) has been effectively eliminated.}
    \label{fig:comb}
\end{figure}

In \textbf{\textit{MRS-N Pipeline}}, we used the \textit{sum merging} method. The wavelengths of the three single exposure spectra are firstly aligned. The probability of cosmic rays falling in the same position is almost zero. Based on this condition, for the three points (e.g. x1, x2, x3) of 3 single exposure spectra at the same wavelength, we firstly select the maximum value (suppose x1=max(x1, x2, x3)). Then the variance of the remaining two values (x2 and x3) can be calculated. When x1 is greater than 3$\times$variance(x2, x3), x1 will be considered to be mainly affected by cosmic rays or the false sharp signal and will be replaced by the mean(x2, x3). Otherwise, there is no cosmic ray or false sharp signal here. The 3 points can be summed directly. After testing, the cosmic rays and false sharp emission lines can be removed effectively with this method. 
The result of an example is shown in the Figure \ref{fig:comb}. As can be seen from Figure \ref{fig:comb}, the false sharp emission lines are removed effectively and the S/N ratio has been significantly improved.
 
\subsection{Wavelength recalibration}

The sky light emission lines, which are mostly from the Meinel rotation-vibration bands of OH \citep{1950ApJ...111..555M},  are ubiquitous in spectra of stars, galaxies or nebulae \citep{1992PASP..104...76O, 1996PASP..108..277O}; The sky light usually exhibits rich emission lines and a weak continuum in the optical spectra. The fainter the observed objects, the more serious the sky light lines contaminate its spectra. Therefore, in nebular spectra, the sky light emission lines appear particularly prominent.  Because of their narrow spectral line widths and fixed central wavelengths, the sky light emission lines can be well used as comparison lines to correct the zero point of spectral wavelength.  

\begin{figure}
    \centering
    \includegraphics[width=\textwidth]{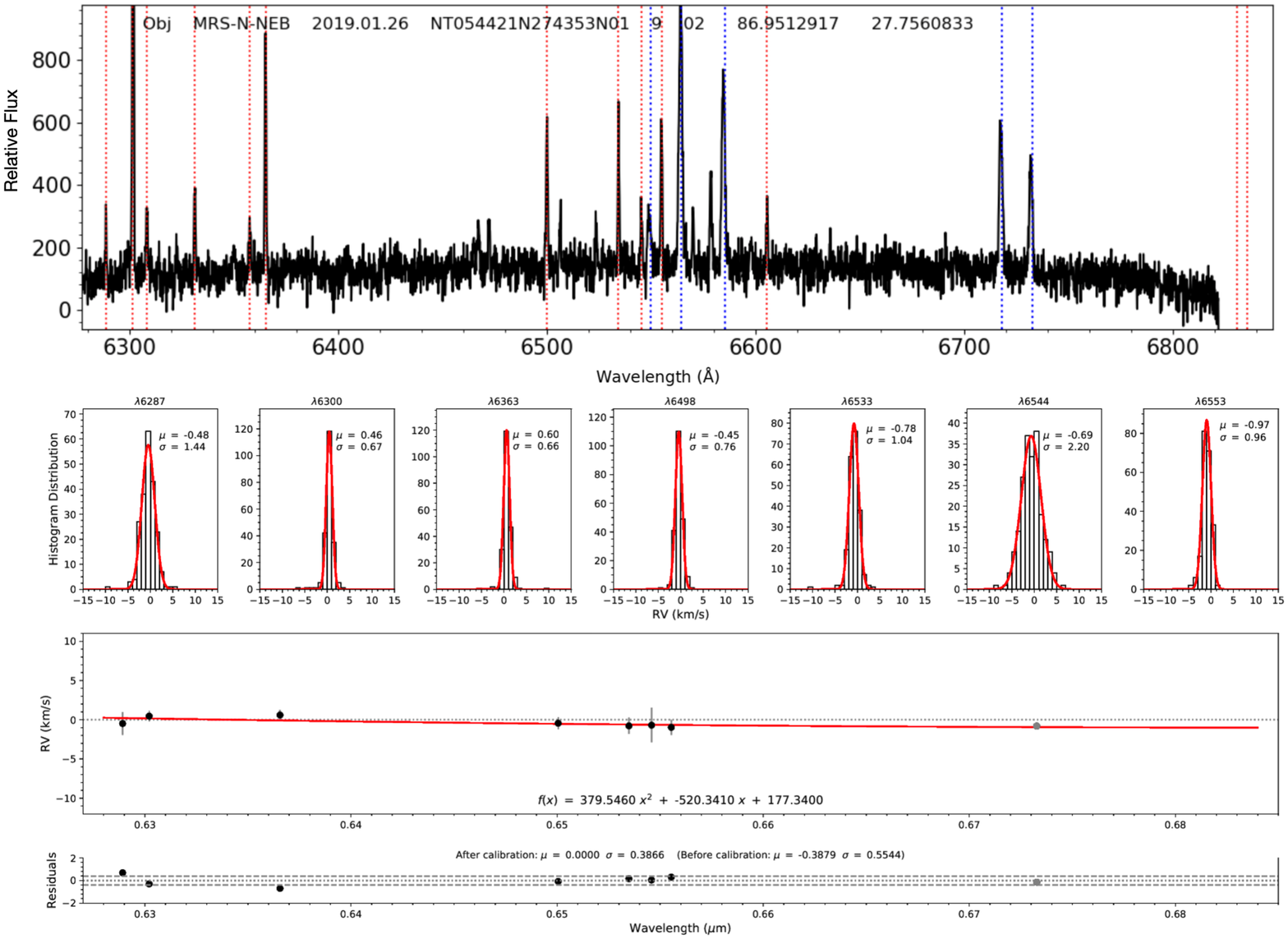}
    \caption{The topmost panel shows an example spectrum (red band) of MRS-N with many skylight lines (red dotted lines) and nebular emission lines (blue dotted lines). Seven of the skylines are fitted with Gaussian function and the fitting results are presented in the middle upper panel. The middle lower panel shows an example of the fitted RV calibration function. The seven black dots represent the fitted mean RVs of the seven single sky emission lines and the gray dot corresponds to the extrapolated value at 6731 $\mathrm{\AA}$. The red curve represents the fitted RVs calibration function. The lowest panel gives the mean values of residuals (gray dotted horizontal line) and the 1$\mathrm{\sigma}$ standard deviations (gray dashed lines).}
    \label{fig:skyline}
\end{figure}

As described in Section \ref{s:pip}, the MRS-N raw data  
is wavelengths calibrated with calibration lamp but without subtracting sky light and correcting flat fields.  
However, the lamp spectra and scientific spectra aren't observed at the same time, which means that the instrument status may have changed during the two observations. Therefore, strictly speaking, it will introduce instrument uncertainties when doing calibration with lamp spectra. This problem can be perfectly solved when the scientific spectra are calibrated with skylight emission lines. Take the red band spectrum of MRS-N as an example, the topmost panel of Figure \ref{fig:skyline} shows a spectrum with the sky light emission lines and nebular emission lines. These sky light emission lines and nebular emission lines are observed at the same time under the same instrument status. Seven of the skylines, $\mathrm{\lambda6287\AA}$, $\mathrm{\lambda6300\AA}$, $\mathrm{\lambda6363\AA}$, $\mathrm{\lambda6498\AA}$, $\mathrm{\lambda6533\AA}$, $\mathrm{\lambda6544\AA}$ and $\mathrm{\lambda6553\AA}$, are fitted with Gaussian function and the fitting results are presented in the middle upper panel of Figure \ref{fig:skyline}. The central wavelengths of the sky light emission lines in scientific spectra should be the same as the theoretical values. So, by fitting the sky light emission lines, 
\cite{2021RAA....21...51R}, which is an important part of the \textbf{\textit{MRS-N Pipeline}}, provided a calibration function ($f(\lambda) = a\lambda^2 + b\lambda +c$, $\lambda$ is the wavelength in unit of $\mu$m, $a$, $b$ and $c$ are the three indexes fitted with sky light lines) by using this method to correct the RVs of scientific spectra in real time. The middle lower panel of Figure \ref{fig:skyline} shows an example of the fitted RV calibration function. The seven black dots represent the fitted mean RVs of the seven single sky emission lines and the gray dot corresponds to the extrapolated value at 6731 $\mathrm{\AA}$. The red curve represents the fitted RVs calibration function. The lowest panel gives the mean values of residuals (gray dotted horizontal line) and the 1$\mathrm{\sigma}$ standard deviations (gray dashed lines). \cite{2021RAA....21...51R} concluded that the systematic deviation can be effectively corrected with this method, especially for the $\mathrm{H\alpha}$ and [\ion{N}{ii}]$\mathrm{\lambda \lambda 6548,6584}$. More detailed can be seen in this reference. 

\vspace{5cm}

\subsection{Subtracting Sky Light}

Sky light subtraction is necessary for the ground-based spectrograph. The traditional method \citep{2016MNRAS.458.3210S} of removing sky light is not suitable for MRS-N data. 
\cite{2021arXiv210808021Z} introduced a method of subtracting sky light for spectra of MRS-N based on the relation of $\mathrm{I(H\alpha_{sky})/I(\lambda 6554)}$ and solar altitude, which is the angle of the sun relative to the Earth’s horizon. $\mathrm{I(\lambda 6554)}$ is the flux of the OH at 6554\AA\,which is from the Earth’s atmosphere. \cite{2021arXiv210808021Z} concluded that $\mathrm{I(H\alpha_{sky})/I(\lambda 6554)}$ and solar altitude have the following relationship:
$$
	\mathrm{I(H\alpha_{sky})/I(\lambda 6554) = 0.954 - 0.011 \times (-Sunalt)}
$$

\begin{figure}
    \centering
    \includegraphics[width=\textwidth]{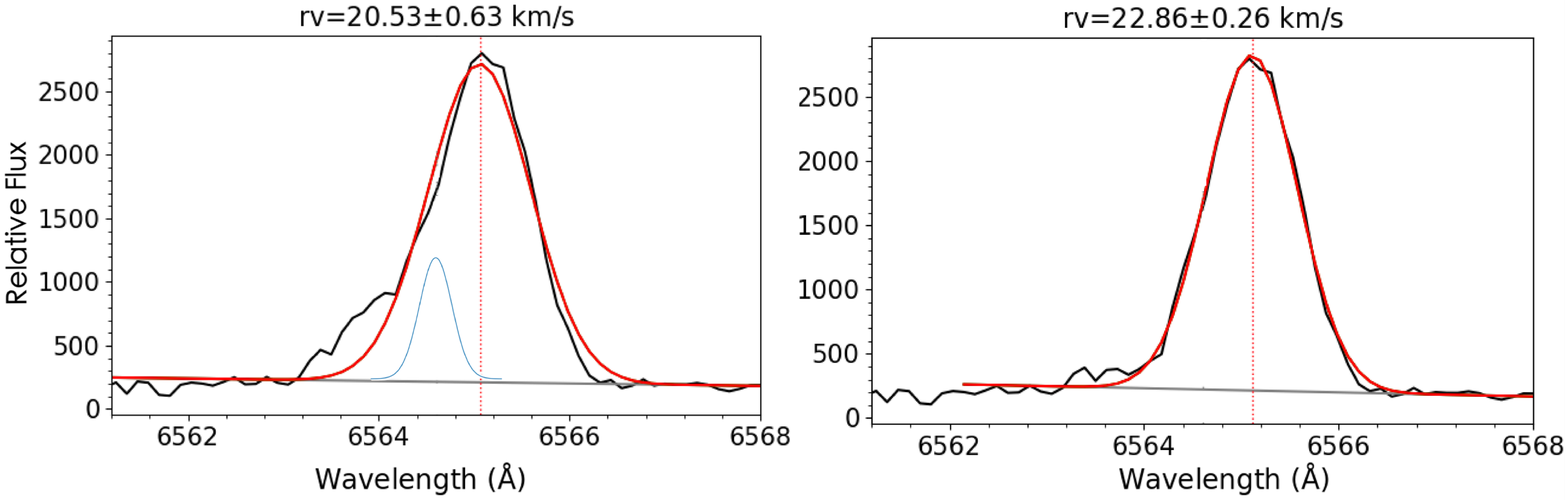}
    \caption{An example of before and after subtracting sky light. \textit{left panel}: Red curve represents the Gaussian fitting of $\mathrm{H\alpha}$ before subtracting sky light, blue curve represents the Gaussian fitting of sky light. \textit{right panel}: Red line represents the Gaussian fitting of $\mathrm{H\alpha}$ after subtracting sky light. The red dotted lines in the two panels show the centroids of nebular $\mathrm{H\alpha}$ emission line.}
    \label{fig:subsky}
\end{figure}

So, once the altitude of the observed target is known, the ratio of $\mathrm{I(H\alpha_{sky})/I(\lambda 6554)}$ can be obtained. We can calculate the $\mathrm{I(\lambda 6554)}$ by single Gaussian fitting. The $\mathrm{I(H\alpha_{sky})}$ can be constructed with three parameters of Gauss function: the line centroid ($\mu$), line dispersion ($\sigma$) and line intensity ($\mathrm{I}$).  Where $\mathrm{\mu_{H\alpha_{sky}} = 6563/6554 \times RV(\lambda 6554)}$, $\sigma_{H\alpha_{sky}} = \sqrt{\sigma(\lambda 6554)^2 + 0.02^2} \mathrm{\AA}$ (0.02 $\mathrm{\AA}$ is an extra broadening caused by the two components of atmospheric $\mathrm{H\alpha}$ emission \citep{2021arXiv210808021Z}), and  $\mathrm{I(H\alpha_{sky}) = ratio \times I(\lambda 6554)}$, where $\mathrm{ratio = 0.954 - 0.011 \times (-Sunalt)}$. Then the constructed $\mathrm{I(H\alpha_{sky})}$ is subtracted from the MRS-N spectra.  By using this method, 90\% of the sky light mixed in nebular $\mathrm{H\alpha}$ emission lines can be reduced, which significantly improves the accuracy of nebulae classification and the measurements of nebular physical parameters. Figure \ref{fig:subsky} shows an example of before (left panel) and after (right panel) subtracting sky light. In the left panel of Figure \ref{fig:subsky}, the red curve represents the Gaussian fitting of $\mathrm{H\alpha}$ before subtracting sky light, blue curve represents the Gaussian fitting of sky light. In the right panel of Figure \ref{fig:subsky}, the red curve represents the Gaussian fitting of $\mathrm{H\alpha}$ after subtracting sky light.

\subsection{Measurements of Nebular Parameters}
\label{s:parm}

\begin{figure}
    \centering
    \includegraphics[width=0.95\textwidth]{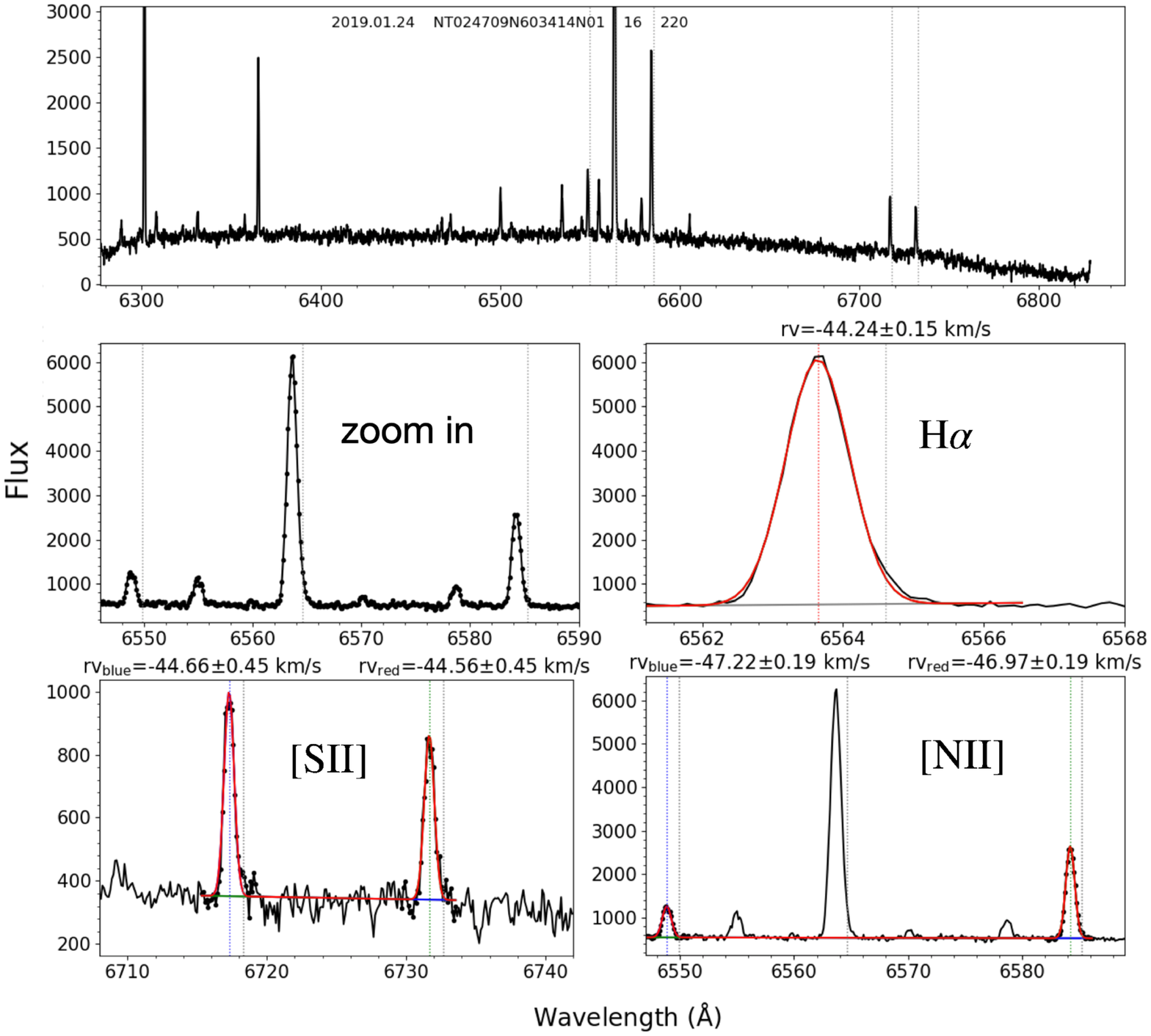}
    \caption{The Gaussian fitting of three nebular emission lines. All the nebular paramers are from the fitting results.}
    \label{fig:fit}
\end{figure}

All the physical parameters, such as RVs, full width at half maximum
 (FWHM), line intensity, etc., are measured by the Gaussian fitting method, which is widely utilized in spectra of the Sloan Digital Sky Survey (SDSS) \citep{2016MNRAS.458.3808R} and LAMOST spectra \citep{2018MNRAS.477.4641R}. We first fit all the science spectra with a second-order polynomial plus a single-Gaussian line profile.  According to the $\chi^2$ and fitting error, the spectra with larger $\chi^2$ value and fitting error will be fitted again by the function of a second-order polynomial plus a double/triple-Gaussian line profile \citep{2021RAA....21...51R}.  Figure \ref{fig:fit}
shows three examples of the single-Gaussian fitting results of $\mathrm{H\alpha}$, [\ion{N}{ii}] and [\ion{S}{ii}] emission lines which are from one MRS-N nebular spectrum.

\vspace{1cm}

\section{Data Products}
\label{s:dp}

Until now, more than 190 thousands nebular spectra have been observed in MRS-N. The data processing is in progress. The data products generated by the \textbf{\textit{MRS-N Pipeline}} contains the following:

\textbf{Nebular Spectra}: The single exposure spectra and coadded spectra are all stored in \textbf{fits} format. They are all wavelength-recalibrated and sky-subtracted spectra. In these spectra, there are no spectra of stars. The stellar spectra observed in MRS-N will be processed by LAMOST 2D and 1D pipeline or LSP3. Since there are fewer nebular emission lines at the blue band of MRS \citep{RAA4768}, these spectral data are mainly for the red band, which covers the wavelength range 6300$\mathrm{\AA}$ - 6800$\mathrm{\AA}$ with $\mathrm{R\sim7500}$. 
The \textbf{fits} files of single exposure spectra are named in the form of \textbf{spec-XXr-PID-YYYYMMDD-N.fits}, where \textbf{XX} represents the spectrograph number (between 01 and 16), \textbf{PID} is the name of observed plate, \textbf{YYYYMMDD} shows the date of observation, \textbf{N} indicates the number (between 01 and 03) of single exposure. The structure of a single exposure spectrum is the same as the structure of LAMOST raw data (See http://dr1.lamost.org/doc/data-production-description\#toc\_3). 

\begin{figure}
    \centering
    \includegraphics[width=0.9\textwidth]{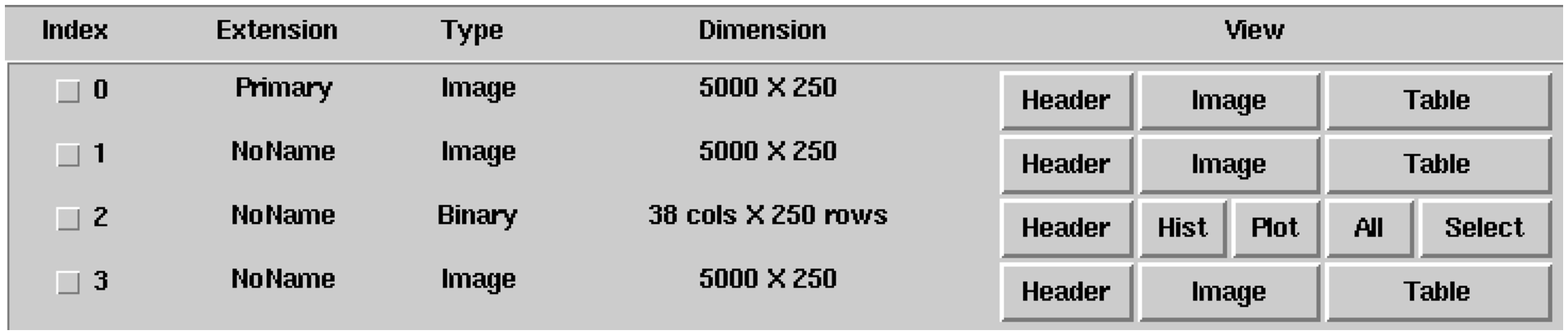}
    \caption{The structure of a coadded spectrum.}
    \label{fig:info}
\end{figure}

The names of coadded spectra are in the form of \textbf{sumspec-XXr-PID-YYYYMMDD.fits}. The string \textbf{XX}, \textbf{PID} and \textbf{YYYYMMDD} have the same meanings as above. Figure~\ref{fig:info} shows the structure of a coadded spectrum. From Figure~\ref{fig:info}, we can see that each fits file has four extensions (EXTEN0, EXTEN1, EXTEN2 and EXTEN3). EXTEN0, EXTEN1 and EXTEN3 represent the relative flux, recalibrated wavelength and invert variance \citep{2015RAA....15.1095L} of EXTEN0, respectively. EXTEN2 shows the information of observed targets. Each extension includes 250 rows, indicating the spectral data of 250 fibers mounted on every spectrograph. 

\vspace{1cm}

\textbf{Nebular Parameters Catalog}: The catalogs of nebular parameters are also stored in \textbf{fits} format. Each catalog contains 41 columns. Table \ref{tab:info} shows the detailed description for each column. 

At present, it is difficult to find a large-scale and multi-target nebular spectra survey to compare with MRS-N in the world. Therefore, we can't do external comparison for all the MRS-N data. It can be considered that there is little difference about the parameters in a certain region (e.g. $< 30^{\prime\prime}$) within a nebula. Based on this condition, we selected the data in \textit{S147} region and compared it with the results of \cite{2018RAA....18..111R}. Moreover, we also made an internal comparison with the data in the same region. 

\begin{figure}
    \centering
    \includegraphics[width=\textwidth]{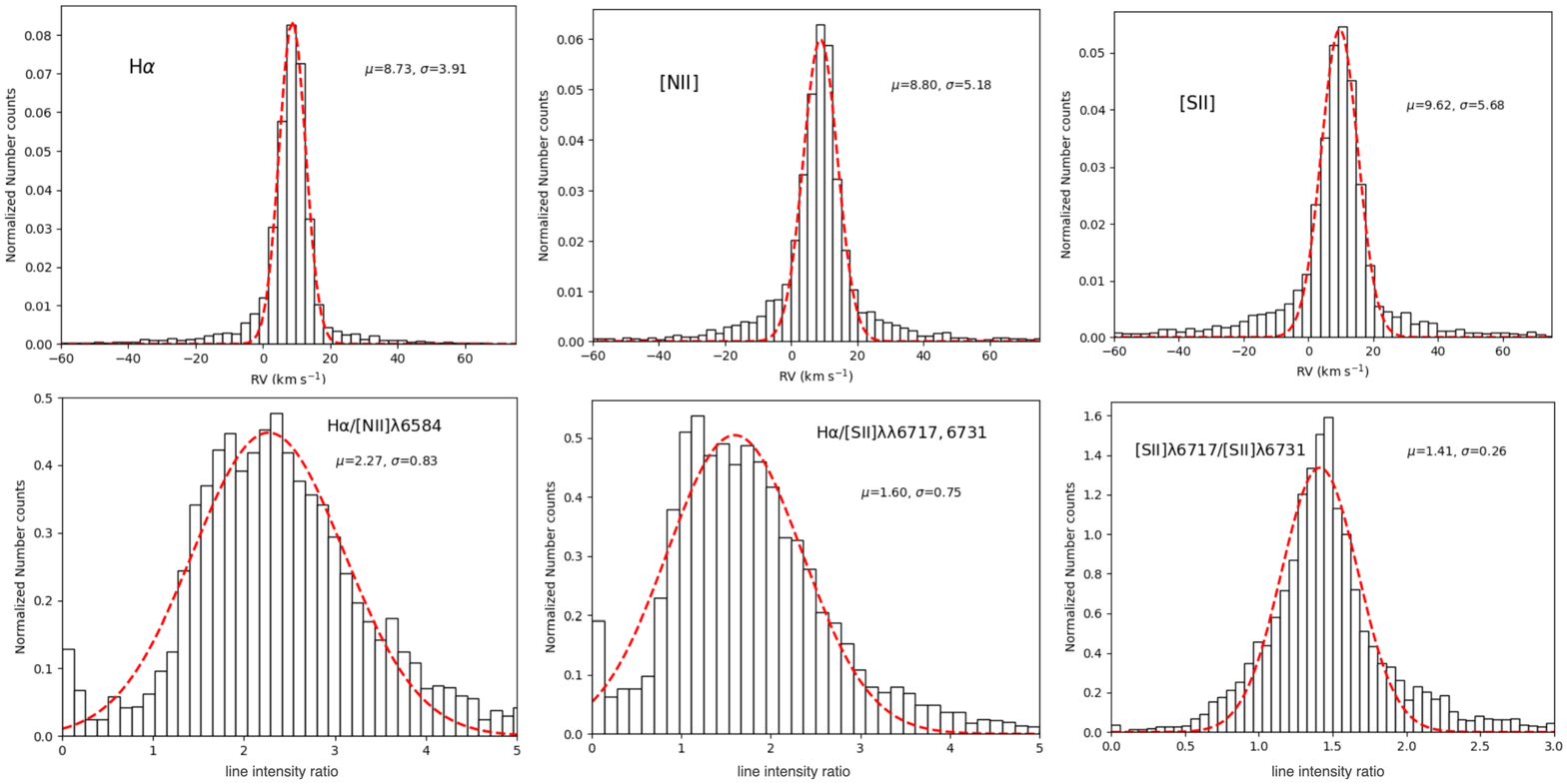}
    \caption{The results of external comparison with \cite{2018RAA....18..111R}.}
    \label{fig:external}
\end{figure}

\textit{External comparison}: In S147 region, by cross matching with the data of \cite{2018RAA....18..111R}, about 480 MRS-N spectra were obtained. We compared the parameters of matched $\sim$480 targets with that of \cite{2018RAA....18..111R}. For this comparison, RVs and line intensity ratio of \cite{2018RAA....18..111R} are from the LAMOST low resolution spectra. Similar to Figure 8 and 10 of \cite{2018RAA....18..111R}, we also gave the histogram distribution of RVs and line intensity ratio (See Figure \ref{fig:external}). The RVs of H$\alpha$, [\ion{N}{ii}] and [\ion{S}{ii}] peak at $\sim$8.73 with $\sigma=3.91$, $\sim$8.80 with $\sigma=5.18$ and $\sim$9.62 with $\sigma=5.68$, respectively. Our results are consistent with the measurements of \cite{2018RAA....18..111R} but with lower dispersion. This is obvious because our spectral resolution is higher. About the line intensity ratios, the peak ($\sim1.41$) of $\mathrm{[\ion{S}{ii}]\lambda6717 / [\ion{S}{ii}]\lambda6731}$ in this work is consistent with that ($\sim1.35$) of \cite{2018RAA....18..111R}. However, the peaks of $\mathrm{H\alpha / [\ion{N}{ii}]\lambda6584}$ and $\mathrm{H\alpha / [\ion{S}{ii}]\lambda\lambda6717,6731}$ in our work are $\sim$2.27 and $\sim$1.60, which are larger than the results given by \cite{2018RAA....18..111R} and with larger dispersion. In fact, they are still consistent within 1$\sigma$ range. One reason of larger dispersion may be that the two methods of subtracting skylight (mainly affects H$\alpha$ emission line) are different; Compared with the method of \cite{2018RAA....18..111R} (selecting the dark area among S147 as the sky light background), our method of subtracting sky light is more reasonable. So we prefer to make a preliminary judgment that the wider dispersion may be caused by the physical factor, but it is not observed in the low resolution spectra. 

\begin{figure}
    \centering
    \includegraphics[width=\textwidth]{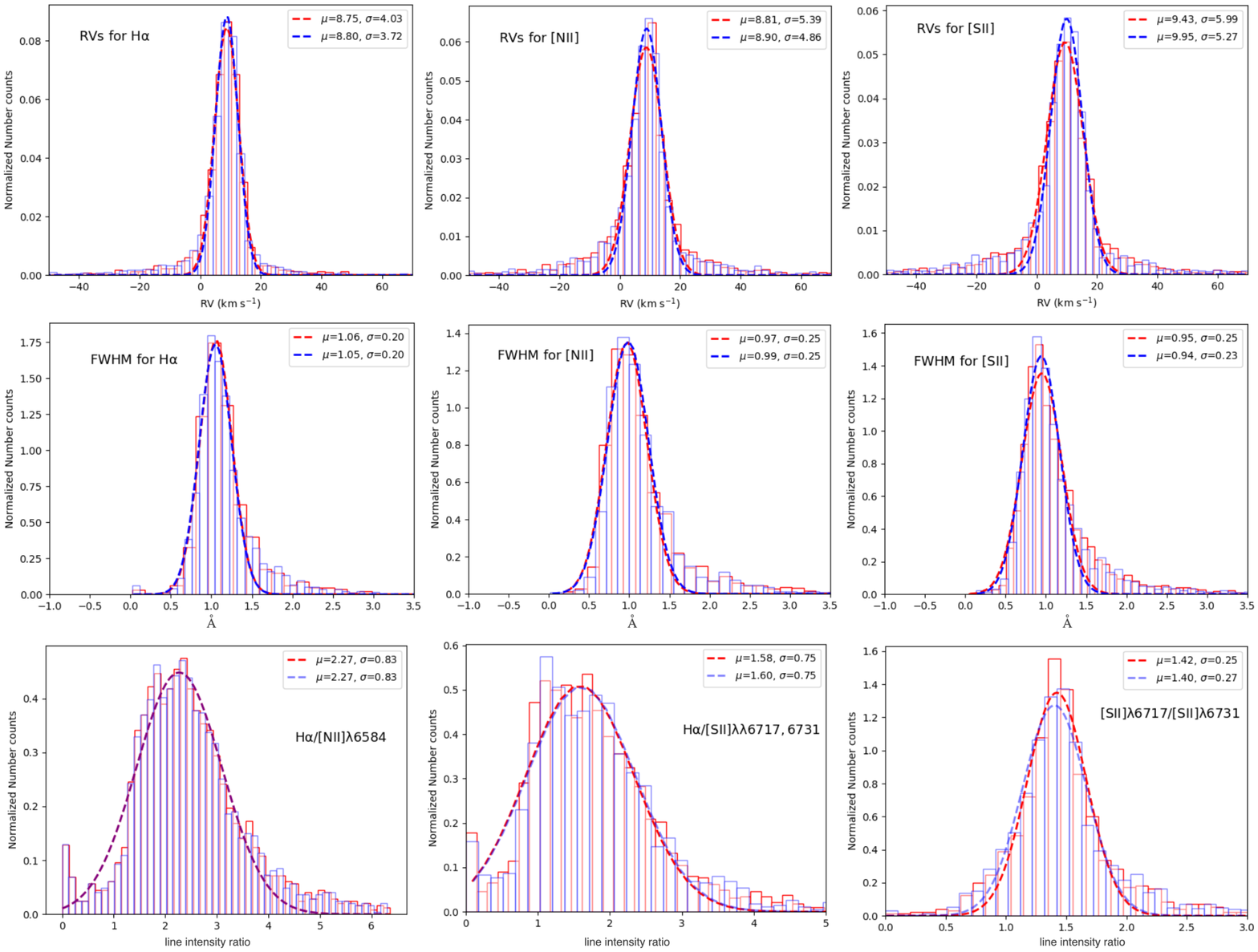}
    \caption{The results of internal comparison within the \textit{S147} region. The spectra in the \textit{S147} region were divided into two groups evenly. The distributions of the two groups were marked with red and blue lines respectively. The red and blue dashed lines represent the fitting results of two groups.} 
    \label{fig:internal}
\end{figure}

\textit{Internal comparison}:  We randomly divided the MRS-N spectra within the \textit{S147} region into two groups (the two groups are evenly distributed in the \textit{S147} region) for comparison. Unlike stellar spectra, the comparison here is for the spectra obtained in the nearby but not the same coordinates. Figure \ref{fig:internal} demonstrates the results of comparison. In Figure \ref{fig:internal}, the top three panels show the distribution of RVs for H$\alpha$, [\ion{N}{ii}] and [\ion{S}{ii}]; the middle three panels represent the distribution of FWHMs for H$\alpha$, [\ion{N}{ii}] and [\ion{S}{ii}]; the bottom three panels give the distribution of line intensity ratio for $\mathrm{H\alpha / [\ion{N}{ii}]\lambda6584}$, $\mathrm{H\alpha / [\ion{S}{ii}]\lambda\lambda6717,6731}$ and $\mathrm{[\ion{S}{ii}]\lambda6717 / [\ion{S}{ii}]\lambda6731}$. The distributions of the two groups were marked with red and blue lines respectively. The red and blue dashed lines represent the Gaussian fitting results of two groups. This comparison result can qualitatively show that the MRS-N results are consistent to a certain extent. 

\section{Summary}
\label{s:summary}
As one of the largest nebular spectra survey on the northern GP, LAMOST MRS-N has conducted for three years (since Oct. 2018) and accumulated more than 190 thousands medium-resolution nebular spectra. In order to make it easier for users to understand and use data, we developed the \textbf{\textit{MRS-N Pipeline}} for the reduction of MRS-N data. The \textbf{\textit{MRS-N Pipeline}} should be used in combination with the LAMOST 2D Pipeline. It mainly includes removing cosmic rays, merging single exposure, fitting sky light emission lines, wavelength recalibration, subtracting skylight, measuring nebular parameters, creating catalogs and packing spectra. In addition, to improve the utilization
of fibers, the fiber without coordinates can also be recalibrated. By using the \textbf{\textit{MRS-N Pipeline}}, the accuracy of nebulae classification and the measurements of nebular physical parameters can be significantly improved.
 
\normalem
\begin{acknowledgements}

    This project is supported by the National Natural Science Foundation of China (Grant Nos. 12073051, 12090041, 12090040, 11733006, 11403061, 11903048, U1631131, 11973060, 12090044, 12073039, 11633009, U1531118), and the Key Laboratory of Optical Astronomy, National Astronomical Observatories, Chinese Academy of Sciences, and the Key Research Program of Frontier Sciences, CAS (Grant No. QYZDY-SSW- SLH007).
    
    C.-H. Hsia acknowledges the supports from the Science and Technology Development Fund, Macau SAR (file No. 0007/2019/A) and Faculty Research Grants of the Macau University of Science and Technology (No. FRG- 19-004-SSI). 

    Guoshoujing Telescope (the Large Sky Area Multi-Object Fiber Spectroscopic Telescope LAMOST) is a National Major Scientific Project built by the Chinese Academy of Sciences. Funding for the project has been provided by the National Development and Reform Commission. LAMOST is operated and managed by the National Astronomical Observatories, Chinese Academy of Sciences

\end{acknowledgements}

\bibliographystyle{raa}
\bibliography{ms}

\clearpage
\begin{center}
\begin{longtable}[p]{ll}
\label{tab:info} \\   
\caption{The description of MRS-N parameters catalog}\\
\noalign{\smallskip}\hline\noalign{\smallskip}
Name of each column & Description  \\
\noalign{\smallskip}\hline\noalign{\smallskip}
obsdate & Date of observation \\
plate   & The name of observed region \\
sp & The number of spectrograph\\
fiber & The number of fiber in each spectrograph\\
fibertype & Fiber type of target, such as Obj, Sky, F-std, Unused, PosErr, Dead\\
objtype & Object type from input catalog, such as Nebula, Skyline, F-star, Star...\\
RA & Fiber pointing Right Ascension (J2000), in degrees\\
DEC & Fiber pointing Declination (J2000), in degrees\\
$\mathrm{SN_{H\alpha}}$ & S/N of $\mathrm{H\alpha}$ \\
$\mathrm{SN_{[N6548]}}$ & S/N of [\ion{N}{ii}] at 6548$\mathrm{\AA}$ \\
$\mathrm{SN_{[N6584]}}$ & S/N of [\ion{N}{ii}] at 6584$\mathrm{\AA}$ \\
$\mathrm{SN_{[S6717]}}$ & S/N of [\ion{S}{ii}] at 6717$\mathrm{\AA}$ \\
$\mathrm{SN_{[S6731]}}$ & S/N of [\ion{S}{ii}] at 6731$\mathrm{\AA}$ \\
$\mathrm{RV_{H\alpha}}$ & Heliocentric RVs of $\mathrm{H\alpha}$\\ 
$\mathrm{errRV_{H\alpha}}$ & Uncertainty of RVs for $\mathrm{H\alpha}$\\
$\mathrm{RV_{[NII]}}$ & Heliocentric RVs of [\ion{N}{ii}]\\ 
$\mathrm{errRV_{[NII]}}$ & Uncertainty of RVs for [\ion{N}{ii}]\\
$\mathrm{RV_{[SII]}}$ & Heliocentric RVs of [\ion{S}{ii}]\\ 
$\mathrm{errRV_{[SII]}}$ & Uncertainty of RVS for [\ion{S}{ii}]\\
$\mathrm{\lambda_{H\alpha}}$ & Fitted line centroid of $\mathrm{H\alpha}$ \\
$\mathrm{err\lambda_{H\alpha}}$ & Uncertainty of fitted line centroid for $\mathrm{H\alpha}$ \\
$\mathrm{\lambda_{[N6548]}}$ & Fitted line centroid of [\ion{N}{ii}] at 6548$\mathrm{\AA}$ \\
$\mathrm{err\lambda_{[N6548]}}$ & Uncertainty of fitted line centroid for [\ion{N}{ii}] at 6548$\mathrm{\AA}$ \\
$\mathrm{\lambda_{[N6584]}}$ & Fitted line centroid of [\ion{N}{ii}] at 6584$\mathrm{\AA}$ \\
$\mathrm{err\lambda_{[N6584]}}$ & Uncertainty of fitted line centroid for [\ion{N}{ii}] at 6584$\mathrm{\AA}$ \\
$\mathrm{\lambda_{[S6717]}}$ & Fitted line centroid of [\ion{S}{ii}] at 6717$\mathrm{\AA}$ \\
$\mathrm{err\lambda_{[S6717]}}$ & Uncertainty of fitted line centroid for [\ion{S}{ii}] at 6717$\mathrm{\AA}$ \\
$\mathrm{\lambda_{[S6731]}}$ & Fitted line centroid of [\ion{S}{ii}] at 6731$\mathrm{\AA}$ \\
$\mathrm{err\lambda_{[S6731]}}$ & Uncertainty of fitted line centroid for [\ion{S}{ii}] at 6731$\mathrm{\AA}$ \\
$\mathrm{F_{H\alpha}}$ & Line intensity of $\mathrm{H\alpha}$ \\
$\mathrm{errF_{H\alpha}}$ & Uncertainty of line intensity for $\mathrm{H\alpha}$ \\
$\mathrm{F_{[N6548]}}$ & Line intensity of [\ion{N}{ii}] at 6548$\mathrm{\AA}$ \\
$\mathrm{errF_{[N6548]}}$ & Uncertainty of line intensity for [\ion{N}{ii}] at 6548$\mathrm{\AA}$ \\
$\mathrm{F_{[N6584]}}$ & Line intensity of [\ion{N}{ii}] at 6584$\mathrm{\AA}$ \\
$\mathrm{errF_{[N6584]}}$ & Uncertainty of line intensity for [\ion{N}{ii}] at 6584$\mathrm{\AA}$ \\
$\mathrm{F_{[S6717]}}$ & Line intensity of [\ion{S}{ii}] at 6717$\mathrm{\AA}$ \\
$\mathrm{errF_{[S6717]}}$ & Uncertainty of line intensity for [\ion{S}{ii}] at 6717$\mathrm{\AA}$ \\
$\mathrm{F_{[S6731]}}$ & Line intensity of [\ion{S}{ii}] at 6731$\mathrm{\AA}$ \\
$\mathrm{errF_{[S6731]}}$ & Uncertainty of line intensity for [\ion{S}{ii}] at 6731$\mathrm{\AA}$ \\
$\mathrm{FWHM_{H\alpha}}$ & Fitted FWHM of $\mathrm{H\alpha}$ \\
$\mathrm{errFWHM_{H\alpha}}$ & Uncertainty of fitted FWHM for $\mathrm{H\alpha}$ \\
$\mathrm{FWHM_{[N6548]}}$ & Fitted FWHM of [\ion{N}{ii}] at 6548$\mathrm{\AA}$ \\
$\mathrm{errFWHM_{[N6548]}}$ & Uncertainty of fitted FWHM for [\ion{N}{ii}] at 6548$\mathrm{\AA}$ \\
$\mathrm{FWHM_{[N6584]}}$ & Fitted FWHM of [\ion{N}{ii}] at 6584$\mathrm{\AA}$ \\ 
$\mathrm{errFWHM_{[N6584]}}$ & Uncertainty of fitted FWHM for [\ion{N}{ii}] at 6584$\mathrm{\AA}$ \\
$\mathrm{FWHM_{[S6717]}}$ & Fitted FWHM of [\ion{S}{ii}] at 6717$\mathrm{\AA}$ \\
$\mathrm{errFWHM_{[S6717]}}$ & Uncertainty of fitted FWHM for [\ion{S}{ii}] at 6717$\mathrm{\AA}$ \\
$\mathrm{FWHM_{[S6731]}}$ & Fitted FWHM of [\ion{S}{ii}] at 6731$\mathrm{\AA}$ \\
$\mathrm{errFWHM_{[S6731]}}$ & Uncertainty of fitted FWHM for [\ion{S}{ii}] at 6731$\mathrm{\AA}$ \\
\noalign{\smallskip}\hline
\end{longtable}
\end{center}

\end{document}